\documentclass[11pt,preprint,graphicx]{aastex}
\usepackage{graphicx}

\def\etal{\it et al. \rm }

\begin{document} 

\title{Stellar Populations and the Star Formation Histories of LSB Galaxies:
III. Stellar Population Models}

\author{James Schombert}
\affil{Department of Physics, University of Oregon, Eugene, OR 97403;
jschombe@uoregon.edu}

\author{Stacy McGaugh}
\affil{Department of Astronomy, Case Western Reserve University, Cleveland, OH 44106;
stacy.mcgaugh@case.edu}

\begin{abstract}

\noindent A series of population models are designed to explore the star formation
history of gas-rich, low surface brightness (LSB) galaxies.  LSB galaxies are unique
in having properties of very blue colors, low H$\alpha$ emission and high gas
fractions that indicated a history of constant star formation (versus the declining
star formation models used for most spirals and irregulars).  The model simulations
use an evolving multi-metallicity composite population that follows a chemical
enrichment scheme based on Milky Way observations.  Color and time sensitive stellar
evolution components (i.e., BHB, TP-AGB and blue straggler stars) are included, and
model colors are extended into the {\it Spitzer} wavelength regions for comparison to
new observations.  In general, LSB galaxies are well matched to the constant star
formation scenario with the variation in color explained by a fourfold
increase/decrease in star formation over the last 0.5 Gyrs (i.e., weak bursts).
Early-type spirals, from the S$^4$G sample, are better fit by a declining star
formation model where star formation has decreased by 40\% in the last 12 Gyrs.

\end{abstract}

\section{Introduction}

The stellar populations in a galaxy is a complicated mixture of stellar
populations with varying ages and chemical composition.  The standard procedure for
untangling unresolved stellar populations in galaxies is to take a population model and
construct a composite spectral energy distribution (SED) from a library of stellar
spectra (Cid Fernandes \etal 2009; Franzetti \etal 2008; Li \& Han 2008; 
Schulz \etal 2002 ) which is then convolved by the filter set of interest and
compared to the photometry of galaxies.  While the evolution of stars is a fairly
well understood field, application of stellar evolution models to galaxies is
complicated by the shear weight of the number of different types of stars involved
(limited by the depth of the stellar spectral library), the difficulty of comparison
observations and complicated ISM physics.

The starting point for a stellar synthesis program is the construction of the
simplest stellar population, one where all the stars are a single age and chemical
composition (e.g., a globular cluster).  As all the details of stellar evolution are
known, then they only variable is the initial mass function (IMF) which assigns the
number of stars per mass bin.  Each mass bin follows a specific evolutionary track
resulting in the color-magnitude diagram as a function of time which is summed to
produce an SED (or a set of colors and indices).  These stellar population models are
called simple stellar populations (SSP's, see Schiavon 2007 for a review) in order to
recognize their limited initial parameters, not to minimize the computational
difficulty of the modeling.

Galaxies, however, are not SSP's simply from the fact that all the stars could not
have been produced in a single, simultaneous burst of star formation.  Star formation
must have been temporally extended, as seen in the metallicity gradients found in
ellipticals and spirals.  Spatially extended information about a galaxy provides a
great deal of information on the changes in age and metallicity internal to the
system but, unless the resolution discerns the actual stars, each resolution element
in the galaxy (i.e., each pixel) is still a composite form of the stellar light.
Thus, the unraveling of the underlying stellar population from its composite spectrum
may be an intractable problem even on small spatial scales.

A great deal of the focus in producing SSP's was on globular cluster and early-type
galaxies due to their assumed simpler histories of star formation.  Star clusters
were a prime target, for the integrated spectrum could be modeled then checked by
direct comparison to the cluster's color-magnitude diagram (CMD, Schiavon 2007).
Ellipticals were the next focus as early color and spectroscopic work indicated their
stellar populations were very similar to globular clusters (Burstein \etal 1984,
Schombert \& Rakos 2009) and they lack any signs of ongoing star formation (i.e.,
H$\alpha$ emission).  Success in modeling ellipticals is measured by the enormous
volume of work using colors and spectral indices to deduce ages and metallicities
(Trager \etal 2000; Kuntschner \etal 2001; Poggianti \etal 2001; Gallazzi \etal 2005;
Thomas \etal 2005; S\'{a}nchez-Bl\'{a}zquez \etal 2006; Rakos, Schombert \& Odell
2008).  While a large number of issues still need to be resolved, such as the details
of short-lived stellar phases (blue horizontal branch and TP-AGB stars) and their
impact on present colors and the evolution of color with redshift, progress in the
details of stellar population synthesis has been substantial (Conroy, Gunn \& White
2009).

Spirals are much more difficult objects to model as their total stellar populations
include an elliptical-like bulge and a star-forming disk (to varying degrees along
the Hubble sequence).  This leads to a problem in interpreting colors and spectral
features due to the age-metallicity degeneracy (Worthey \etal 1995) and the
complications due to a highly variable (spatially and in density) ISM (Kennicutt
\etal 2009).  Irregulars were actually better candidates to study stellar populations
due to their closeness to the Milky Way and more visible star formation regions (less
optical depth), as well as typically lower stellar masses and metallicities which
serves to compress the age-metallicity degeneracy effect (Hunter \& Elmegreen 2004).

Successful modeling of spirals and irregulars has reproduced many of their large
scale characteristics, such as the color-magnitude and the Tully-Fisher relation
(Boissier \& Prantzos 2000, Bell \& Bower 2000).  These models have also been
successful at understanding the relationship between gas density and the star
formation rate (SFR), used to develop various star formation laws (Kennicutt \& Evans
2012).  Adding a chemical evolution model allowed for predictions for the 
mass-to-light ratios ($M/L$) and the various color relationships for spirals (Bell \&
de Jong 2000).  Almost all of the bright, high surface brightness spirals are well
fit with models of declining SFR's, matching the loss of their gas supply with time
(Boissier \& Prantzos 2000).

Low surface brightness (LSB) galaxies are, in general, not particularly small or low
in mass (see Schombert, Maciel \& McGaugh 2011).  Their only defining characteristic
are central surface brightnesses below 23 B mag arcsecs$^{-2}$ (the typical spiral
has a central surface brightness of 21 B mag arcsecs$^{-2}$).
A dilemma is posed by LSB galaxies, for their gas fractions
are high, their stellar densities are low, their optical colors are blue and their
ratio of stellar mass to mean SFR implies a timescale of star formation of a Hubble
time (McGaugh \& de Blok 1997).  The combination of these factors indicates a star
formation history that has a much lower mean SFR than most spirals and irregulars
plus a history that must be nearly constant SFR (or perhaps evenly spaced bursts),
since the time of galaxy formation in order to produce enough stellar mass, yet have
present day optical colors that are extremely blue.

While appearing to represent a challenge to star formation models, LSB galaxies may
also provide a clearer window into the star formation history of gas-rich galaxies for
several reasons.  One, they contain much less dust than normal spirals or irregulars
(O'Neil \& Schinnerer 2003).  While they are rich in neutral hydrogen, their low
stellar surface densities also reflect into low gas cross sections (McGaugh \& de
Blok 1997), so the effects of the ISM (both dust and gas) on colors are minimal.
Second, all indications are that LSB galaxies have very low mean metallicities
(McGaugh 1994).  This reduces the range of population models that must be integrated to
produce a total spectrum or color and reduces the complications from the different
tracks for metal-rich versus metal-poor stars.  Third, since the ratio of SFR to
stellar mass in LSB is close to unity for a Hubble time then the mean SFR has been
nearly constant for LSB galaxies, as a class, over their evolutionary history
(Schombert, McGaugh \& Maciel 2013, Schombert \& McGaugh 2014) which greatly
simplifies the modeling.  The duration and strength of burst activity can be
constrained by examining the scatter in the H$\alpha$ luminosity versus galaxy mass
relation.  All the above dramatically reduces the possible model histories down to a
set of quasi-static constant SFR scenarios.

The goal of this paper is to outline a series of models to explain the colors of LSB
galaxies, using the above constraints on the properties of LSB galaxies that enhance
the reliability of population synthesis.  Even with the scenario of constant star
formation, there still a series of unknowns that must be explored in the models to
reproduce the full range of possible color histories.  Our models are also extended
to wavelengths within the {\it Spitzer} telescope imaging range, providing longer
wavelength leverage which is critical to separate the influence of TP-AGB stars,
metallicity effects and explore a realm where $M/L$ estimates are more stable
(McGaugh \& Schombert 2014).

\section{Model Parameters}

\subsection{Single Age Composite SSP's (multi-metallicity models)}

The obvious next step beyond a SSP, a population of single age and single
metallicity, is a single age composite SSP which includes a range of metallicities
that follow a particular enrichment scenario (Pagel 1997).  Of course, physically, it
is impossible to generate enriched stars without a progression of stellar birth and
death, then recycling enriched material back to the ISM.  This clearly takes a finite
amount of time but, fortunately, the recycling timescales are much shorter than
noticeable changes in a stellar population due to age (Matteucci 2007).  Enrichment
can be treated as instantaneous and the stars can be assumed to arise from an
enriched gas with a range of gas metallicities that reflect the metallicity
distribution function (MDF) of the stars.

With a proper MDF, a composite stellar population is constructed by extracting
from the stellar library stars of a single age and varying metallicities as given by
the MDF bins.  The resulting spectra are summed and weighted by number to produce
integrated colors and indices.  The mean metallicity is calculated either as a
luminosity weighted value, or a averaged value (i.e. by number).  The remaining
problem is to determine a MDF that fits the stellar populations in a star-forming
galaxy.

\begin{figure}[!ht]
\centering
\includegraphics[scale=0.80,angle=0]{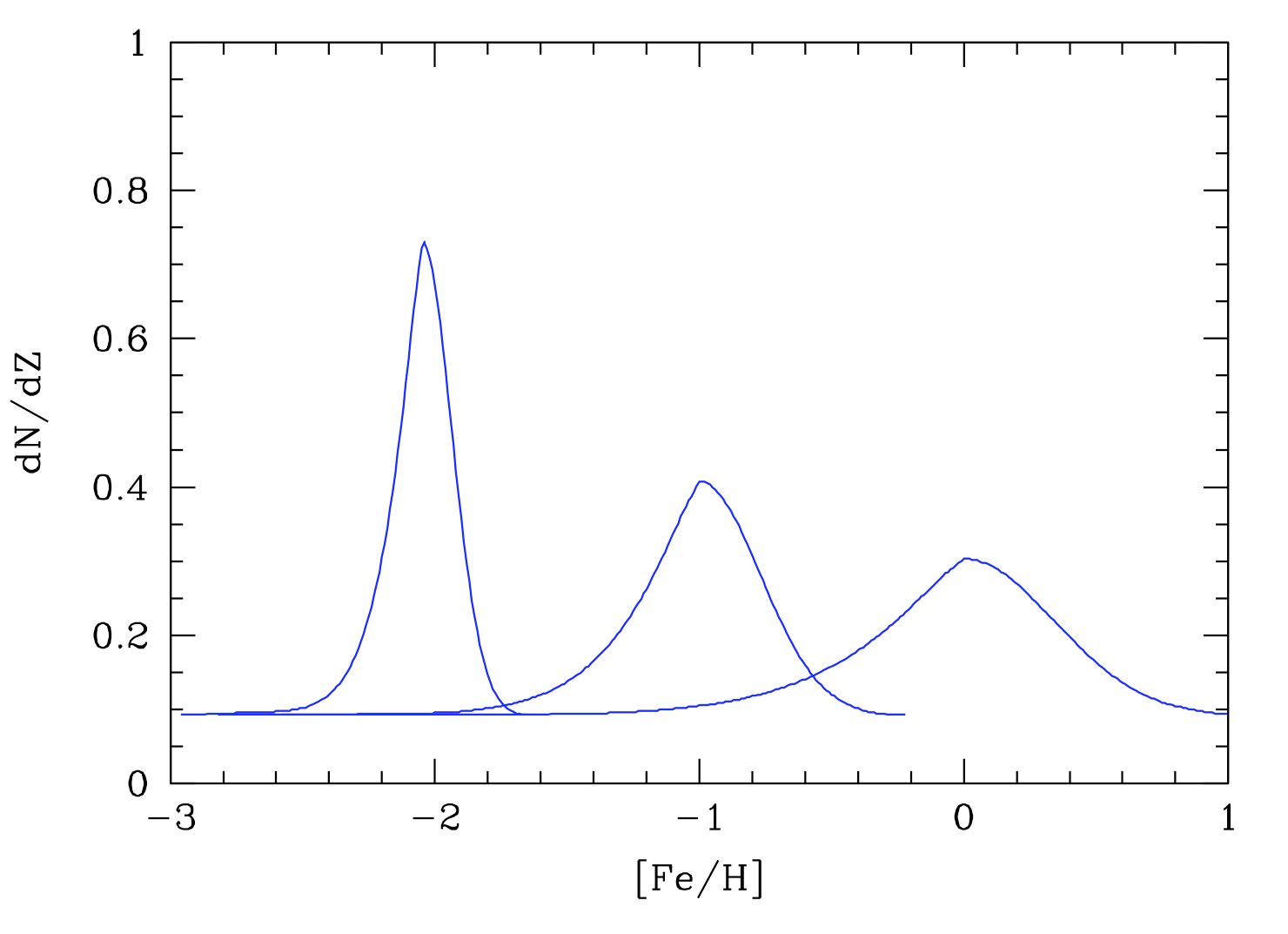}
\caption{\small Three `push' MDF's are shown for peak [Fe/H] of $-$2, $-$1 and solar,
all having a long tail to low metallicities and a steeper high metallicity side.  The
effect of depressing the decreasing low metallicity stars (the push to resolve the
G-dwarf problem) is evident by the shallower slope on the low metallicity side.
Each population is normalized to the same total number, which results in a wider
spread in [Fe/H] with increasing peak [Fe/H].  For the low metallicities, the small
width means these models are identical to a gaussian or any simple infall model.
}
\label{push}
\end{figure}

For the current suite of models, we have followed the MDF prescription outlined in
Schombert \& Rakos (2009) which has the advantage of matching the chemical evolution
of Milky Way stars and correctly reproducing elliptical colors.  The MDF model
adopted in Schombert \& Rakos (2009, called the `push' model) is designed to resolve
the G-dwarf problem (Gibson \& Matteucci 1997, an underestimation of low metallicity
stars in closed box models) and incorporates a mixing infall scenario (Kodama \&
Arimoto 1997, linking the accretion rate to the star formation rate).  The resulting
MDF (see Figure 4, Schombert \& Rakos 2009) provides a simple analytic function that
can be adjusted for its peak [Fe/H] to reproduce any mean [Fe/H] for a stellar
population.

Our `push' model MDF is shown in Figure \ref{push} for three different peak [Fe/H]'s
and its general shape follows MDF's given by theoretical chemical evolution
simulations, nearby star counts and HST studies on the tip of the red giant branch
(RGB) in nearby ellipticals (Harris \& Harris 2000).  Nearby galaxies display a lower
metallicity peak with increasing radius from the galaxy center (i.e., metallicity
gradients).  This results in a narrower MDF with lower peak metallicities, but
maintains the same general shape.  

The success, and usefulness, of this model is tested by the accuracy that the
composite SSP's colors and indices match globular cluster and elliptical colors over
a range of galaxy masses (i.e., total metallicities, Schombert \& Rakos 2009b).  And,
while successful with single burst populations that dominate elliptical stellar
populations, there are elements to the `push' model that make it as useful for
constant star formation scenarios.  For example, the `push' model matches the shape
of inhomogeneous enrichment models (Malinie \etal 1993, Oey 2000) where star
formation occurs in discrete patches throughout a galaxy and is only allowed to mix
between star formation episodes. This increases the amount of mixing and results in
fewer metal-poor stars.  In fact, our model at solar metallicities is identical to
Prantzos (2009) infall model with a timescale of 7 Gyrs.

The use of a realistic metallicity spread has an increasing effect on a stellar
populations colors with increasing mean metallicity.  For a 12 Gyr population with a
mean metallicity of $<$Fe/H$>$ = $-2$, the change in optical and near-IR colors is
effectively zero as the spread in metallicity is very small.  By a mean metallicity
value of $<$Fe/H$>$ = $-1$, the change in $B-V$ and $V-K$ is $-$0.01.  For a solar
metallicity population, the difference is $-$0.05 in $B-V$ and $-$0.02 in $V-K$
(Schombert \& Rakos 2009).  For a younger population, this difference is the same in
the optical (age dominates over metallicity) but increases in the near-IR to
$\Delta(V-K)=-0.09$ for a solar metallicity population.  The effect, overall, is to
make all the colors slightly bluer due to the inclusion of a metal-poor component at
each epoch, an expected result.

\begin{figure}[!ht]
\centering
\includegraphics[scale=0.80,angle=0]{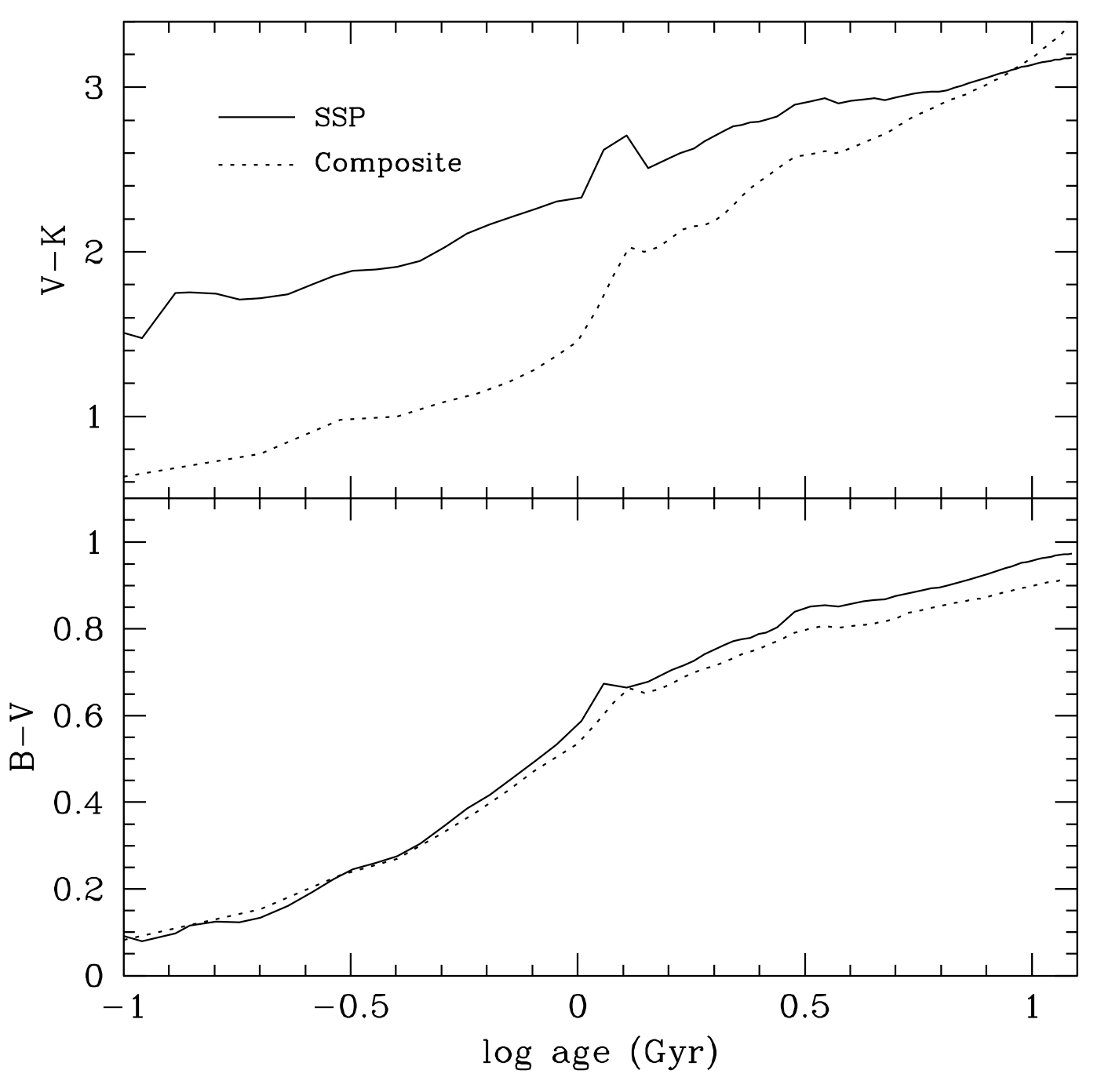}
\caption{\small The effect on $B-V$ and $V-K$ between a single metallicity
population (SSP) and a composite population (using our `push' MDF) as a function of age.
A peak solar metallicity model is displayed and the 
differences in colors from adding a distribution of metallicities is particularly noticeable at
young ages with the combined effects of young age and a contribution from metal-poor
stars drives the RGB to higher temperatures and bluer $V-K$ colors.  The difference
in $V-K$ fades with time, but the metal-poor stars begin to show differences in
optical colors ($B-V$) as the turnoff point reddens after 1 Gyr.  The crossover in
$V-K$ color at 10 Gyrs is due to the shape of the MDF having more stars above solar
[Fe/H] than below.
}
\label{composite_color}
\end{figure}

The effect of using composite metallicity models on galaxy colors varies with
wavelength.  For both near-UV and optical colors, a spread in [Fe/H] results in
slightly bluer colors.  But, for optical colors, the separation in color between
SSP's and composite models increases with increasing stellar population age.  The
bottom panel in Figure \ref{composite_color} displays the run of $B-V$ for a solar
metallicity population as a function of age.  For ages less than 1 Gyr, the
differences are minor.  By 10 Gyrs, the different between the models is 0.05 mags.
While this is small in an absolute sense, it is noticeable in studies of the
color-magnitude diagram in ellipticals (Rakos \& Schombert 2009).

The color differences between the composite metallicity models and SSP's are more
direct for near-IR colors.  The top panel in Figure \ref{composite_color} displays
the run of $V-K$ for the range of ages from 0.1 to 12 Gyrs.  As RGB stars dominate
the near-IR colors in galaxies, and the temperature of RGB branch is strongly
dependent on [Fe/H], strong color differences are found at young ages (0.8 at 0.1
Gyrs) decreasing to zero at 10 Gyrs.  The composite models become redder than the
SSP's after 10 Gyrs due to the fact that a spread of [Fe/H] is involved in the
composite models which finds the metal-rich stars being brighter than the metal-poor
stars for old stellar populations.

The only remaining variable to our MDF simulations is the effect of infalling gas.
LSB galaxies are particular gas-rich and, while it is assumed that the low SFR in LSB
galaxies is due to low gas densities, the total gas mass of these systems is high.
This leads to the increased probability that a significant fraction of the gas supply
for star formation is from primordial infalling gas.  Mixing models account for some
primordial gas in their simulations, but LSB galaxies may have a higher percentage of
metal-poor gas than other spirals or irregulars.  Our only argument against
increasing the metal-poor side of our baseline MDF is that LSB galaxies do not
deviate from the color-magnitude diagram from other star-forming galaxies (Schombert
\& McGaugh 2014).  The color-magnitude relation is driven by a mixture of star
formation, age and metallicity effects (Tojeiro \etal 2013) and the colors of LSB
galaxies place them at bluer colors than normal spirals in line with their lower
metallicities.  There is no indication that their metallicities are extremely low, so
we have not altered our baseline MDF to allow for a much larger population of
metal-poor stars than already accounted for by inhomogeneous mixing scenarios.
We also note that due to the lack of detection of dust in any LSB galaxies, either
due to low metallicities or simple low stellar densities, we have ignored any
internal extinction corrections to our model colors.

\subsection{Chemical Enrichment Model}

For star forming galaxies, such as spirals and irregulars, the MDF will continue to
evolve as long as star formation continues.  The MDF evolution is such that the peak
metallicity increases with time plus the rate of evolution is proportional to
the rate of star formation.  To characterize a changing mean metallicity, the peak
[Fe/H] of our `push' model is simply shifted to higher metallicities while maintaining
its shape and tying the lowest metallicity end to a zero metallicity gas (effectively
$[Fe/H]=-2.5$, see Figure \ref{push}).  Each curve is parameterized by the peak
[Fe/H], but the mean metallicity ($<$Fe/H$>$) and the luminosity weighted [Fe/H] value
vary in a linear fashion (see Schombert \& Rakos 2009).

In order to develop a realistic chemical enrichment scenario, a simple model of
chemical evolution is required constrained by actual age-metallicity relations in
nearby galaxies.  Numerous age-metallicity models have been proposed (Sellwood \&
Binney 2002, Prantzos 2009) all having the similar initial conditions of a $[Fe/H] =
-1.5$ enriched gas then leveling off after 5 Gyrs (the final [Fe/H] dependent on the
mass of the galaxy).  However, most models fail to match the run of [Fe/H] with age
in our own Galaxy (e.g., Nordstrom \etal 2012) where extremely metal-poor stars are
rare and the models over-predict their numbers compared to observations.

The difference between models and observations has been attributed to two effects; 1)
the large statistical uncertainties in the data, particular for the metal-poor end of
the data samples and 2) the impact of radial mixing on the observed Galaxy sample
(Prantzos 2009).  The initial metal building phase is critically important to our
models since most LSB galaxies appear to have total metallicities much less than
normal spirals (McGaugh 1994).  Thus, we need SF models where the final [Fe/H] is in
the range of $-$1.5 to solar and such that the oldest stellar populations with the
lowest metallicities can be included in the synthesis calculations.

\begin{figure}[!ht]
\centering
\includegraphics[scale=0.80,angle=0]{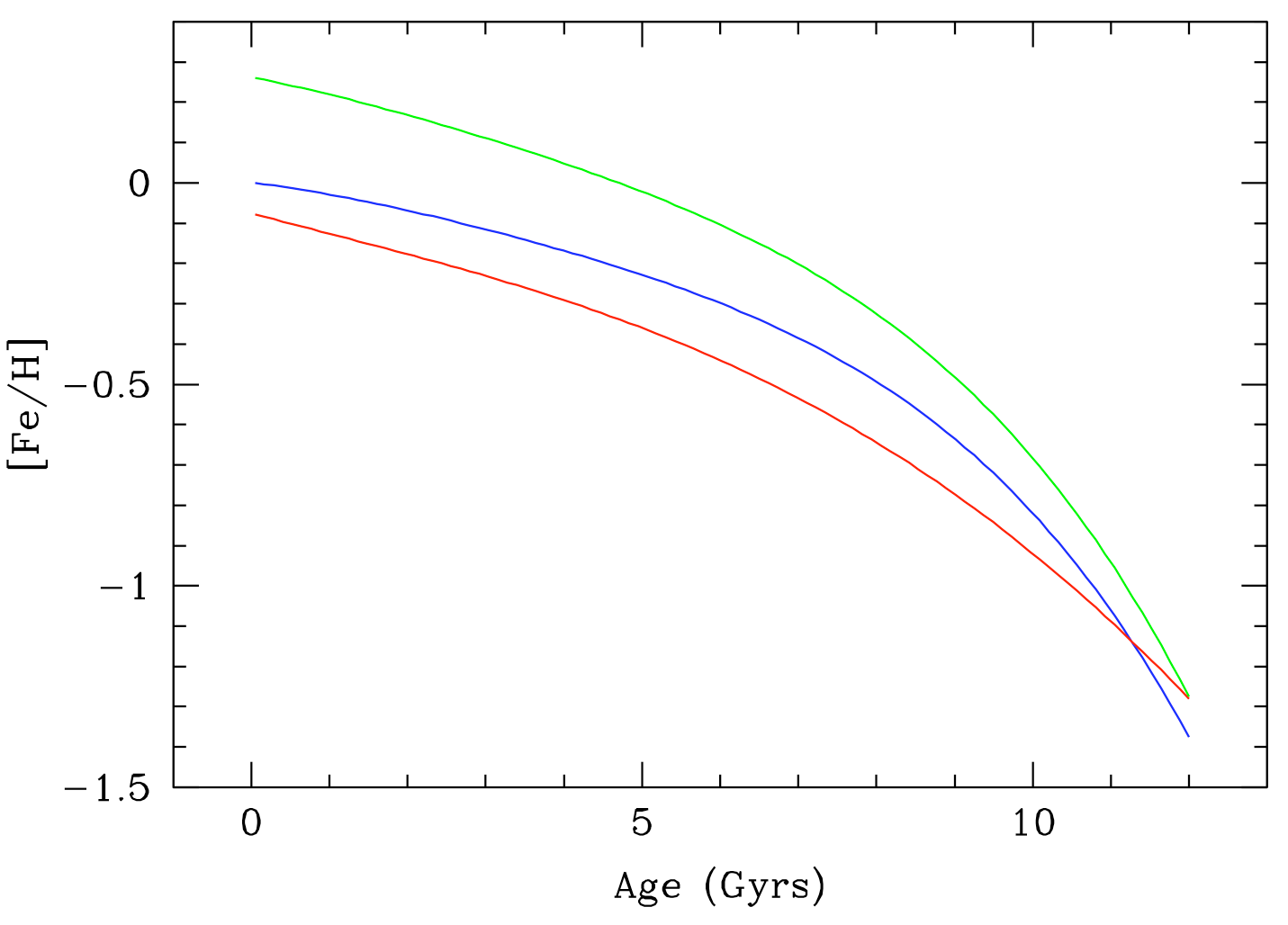}
\caption{\small The chemical enrichment scenario taken from Prantzos (2009) shown as
the blue line.  Fixed at $[Fe/H]=-1.5$ at 12 Gyrs (model start time) and rising to
solar at the present day.  The green line displays the instantaneous [Fe/H] as a
function of age.  Since the `push' model is asymmetric, such that there are more higher
than peak metallicity stars than below the peak, the instantaneous [Fe/H] is higher that
the peak value at all times.  The red line displays the running average metallicity
($<Fe/H>$), displaying the integrated metallicity of all the past stars at each
generation.  While the Prantzos model is used to set the normalization of [Fe/H] with
time, the actual values vary due to the asymmetric shape of the MDF.  The final
averaged value is used for the analysis diagrams in \S4.
}
\label{prantzos}
\end{figure}

To this end, we have adopted the Prantzos enrichment model which rapidly increases
the mean [Fe/H] for the first 2 Gyrs (80\% yield), then an additional 15\% over the
next 5 Gyrs and effectively constant for the last 5 Gyrs (assuming a 12 Gyr old
galaxy, see Figure \ref{prantzos}).  The parameterization for this scenario will be
the final generation metallicity ranging from [Fe/H]=$-$1.5 to 0.2.  To maintain the
same shape, we adjust the shape of the age-metallicity relation so that same
percentages in each metallicity bin are the same from model to model.  The [Fe/H]
value at a particular age is taken to be the peak [Fe/H] for the `push' MDF, thus,
even in the final generations of stars, some metal-poor component is produced to
represent the infall of low metallicity gas into the galaxy.  Using a MDF in
combination with an enrichment model will result in averaged [Fe/H] that are
consistently below the model value (due to the sum of previous metal-poor
populations, see Figure \ref{prantzos}).

The effects on galaxy colors of using a multi-metallicity with a chemical enrichment
scheme are fairly significant.  Compared to a constant SF scenario with an
non-evolving solar metallicity population, a chemical enrichment model produces
little change in $B-V$ colors, they are bluer by 0.05 mags.  However, the difference
in the near-IR is substantial with $\Delta(V-K)$ of 0.15 with a simple chemical
evolution scenario due to the higher sensitive of RGB stars to metallicity.  The
inclusion of a metal-poor component to the composite population has the strongest
effect on the width and position of the composite RGB.

\begin{figure}[!ht]
\centering
\includegraphics[scale=0.80,angle=0]{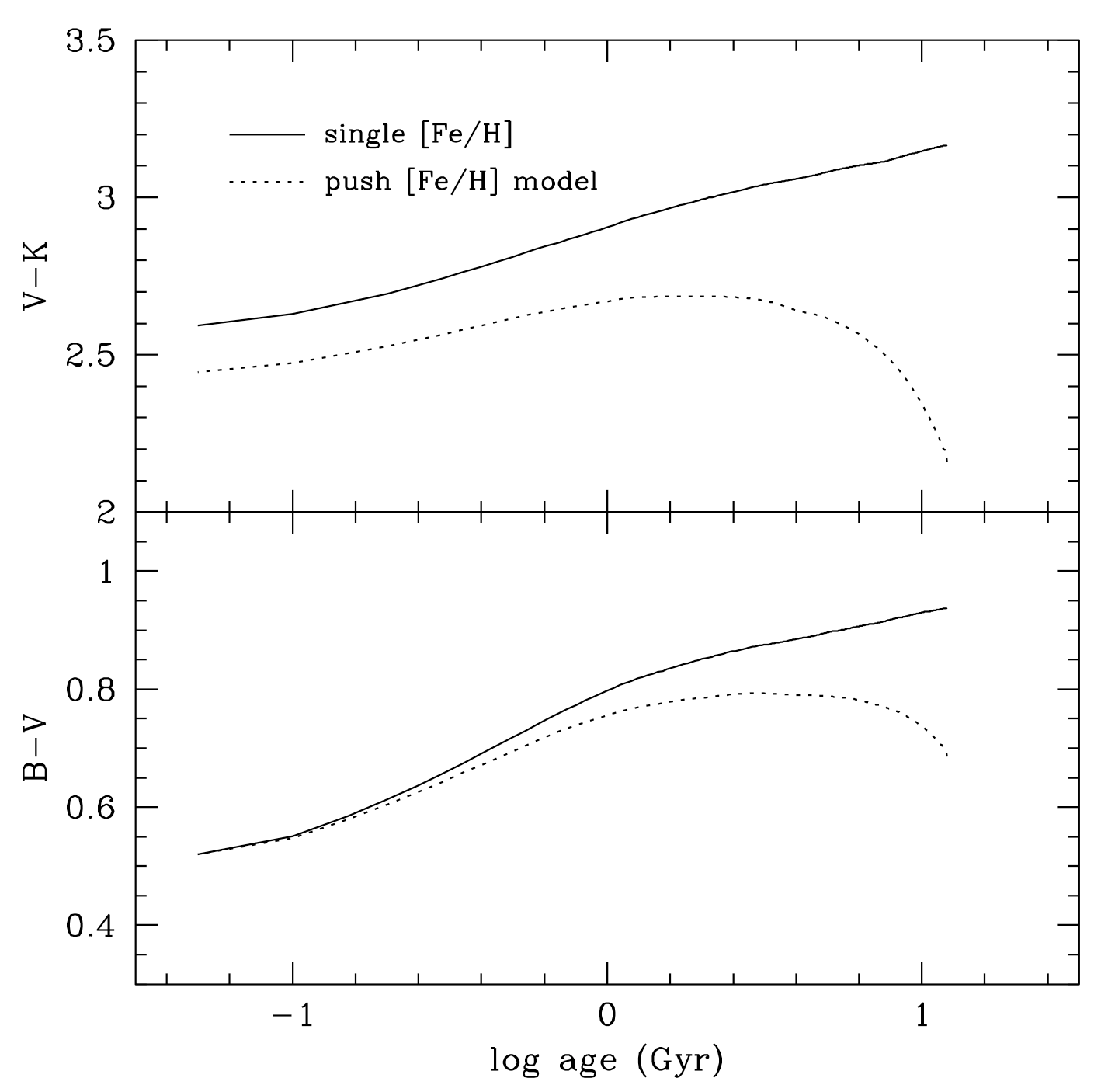}
\caption{\small The difference in $B-V$ (bottom panel) and $V-K$ (top panel) color
for the single metallicity and the composite metallicity models.  The push model
represents a model which starts at $[Fe/H]=-1.5$ ending at solar following the curves
in Figure \ref{prantzos}.  The large colors differences for the older ages is
artificial in the sense that the push colors are for a much lower metallicity
population and, thus, bluer colors.  However, for the younger ages, the optical
colors are dominated by young stars and the metal-poor portion of the underlying
population has no effect.  For near-IR colors, the age effects are less important
compared to the temperature of the RGB, which has a strong contribution from
metal-poor stars in the push model.  
}
\label{chemev_color}
\end{figure}

\subsection{Initial Mass Function}

The distribution of stellar masses is known as the stellar initial mass function
(IMF), i.e. the percentage of stars within specific bins of mass assumed for the
stellar population at the onset of star formation.  While the onset of thermonuclear
fusion may vary by mass within a molecular cloud, the timescale from protostar
formation to the main sequence is shorter than the timesteps assumed by our model (10
Myrs).  Thus, each stellar population segment is assumed to produce a range of masses
already on the main sequence at time zero (i.e., no T-Tauri phase is used in the
simulation).

Observations on the shape of the IMF in other galaxies are very limited (Kroupa 2001)
and mostly revolve around indirect measurements, e.g., the effect on galaxy colors
due to changes in a model IMF (van Dokkum 2008).  For simulations of galaxy colors,
changes in the upper end of the IMF dominate integrated colors due to the high
luminosities of high mass stars.  As a stellar population evolves, the upper end
of the main sequence feeds into the RGB luminosity to effect near-IR colors.  
Changes in the IMF below 1$M_{\sun}$ have a negligible effect on colors, however,
the small changes in the numbers on the lower end of the IMF can have a significant
effect on calculated $M/L$'s for the population (see \S4.3).

For LSB galaxies, we have a mild constraint on the upper end of the IMF from
comparison of star cluster luminosities to H$\alpha$ emission (see Figure 13 of
Schombert, McGaugh \& Maciel 2013).  To match the upper envelope of the cluster mass
to H$\alpha$ luminosity required a power-law index of $-$2.6, compared to the value
of $-$2.5 from the Kroupa \etal (2013) formulation.  A slightly different slope on the
upper end of the IMF has a small effect on the $B-V$ colors (bluer by 0.01), but no
effect on colors redward of $V$.  Most of uncertainly that derives from variation in
the IMF occurs in colors for populations younger than 3 Gyrs, which make
up less than 20\% of the final stellar mass.  Color effects were concentrated in
filters blueward of $V$, although calculations of $M/L$ are sensitive to the IMF even
out to 3.6m$\mu$ (McGaugh \& Schombert 2014).

As noted by Conroy, Gunn \& White (2009), a discontinuous IMF, such as proposed by
Kroupa (2001), has the unfortunate effect of introducing irregular jumps in
luminosity and color at the timescales where the IMF slope changes.  However, the IMF
proposed by Chabrier (2003) is very similar to the Kroupa IMF, at least for masses
less than 10 $M_{\sun}$.  Thus, we have adopted the Chabrier formulation for our
model IMF, with a slightly steeper slope for stars greater than 10$M_{\sun}$
($-$2.6).

\subsection{$\alpha$/Fe Corrections}

The colors of stellar populations is determined by the distribution of stars, given
by stellar isochrones, in the HR diagram.  Decreases in metallicity drive both the
turnoff point and the position of the RGB to hotter temperatures, i.e. bluer colors,
due to changes in line blanketing and opacity.  While Fe is the main contributor of
electrons to produce color changes, all atoms heavier than He can contribute
electrons.  It is typically assumed that all the elements track Fe abundance, but it
is possible to have overabundances of various light (so-called $\alpha$ elements)
under certain conditions.

The elements lighter than Fe (the so-called $\alpha$ elements) are primarily produced
in massive stars (Type II SN), while a substantial contribution to Fe comes from Type
Ia SN.  Since SNII are short-lived ($\tau < 10$ Myrs) and SNIa detonate only after a
Gyr (the mean time for the white dwarf to form), the ratio of $\alpha$/Fe measures
the formation timescales.  At early epochs, the ratio of $\alpha$/Fe is determined by
Type II supernovae (SNII) and after a Gyr, the $\alpha$/Fe ratio decreases due to the
products from SNIa explosions (cf.  McWilliam 1997, and references therein).

For example, in elliptical galaxies the ratio of $\alpha$/Fe is a factor of four
higher than metal-rich stars in the Milky Way due to the short timescales of initial
burst star formation (Thomas \etal 2004).  As star formation is extended in spirals and
irregulars, those galaxy types have lower $\alpha$/Fe ratios as more recent star
formation is enriched by SNIa Fe (Matteucci 2007).  For a constant star formation
scenario, we can model the ratio of $\alpha$/Fe as a function of [Fe/H] following
data from the Milky Way (Milone, Sansom \& Sanchez-Blazquez 2010) with [Fe/H] serving
as a proxy for age.  In their data, elliptical-like $\alpha$/Fe ratios are found up
to $[Fe/H]=-1.0$, then drops quickly to a solar value at solar metallicities.  We
will apply the same behavior to our simulations.

The effect of the $\alpha$/Fe ratio on colors was outlined in Cassisi \etal (2004).
With respect to colors, $B-V$ decreases (bluer) with increasing $\alpha/Fe$, for example a
metal-poor population ([Fe/H]=$-$1.3) had a blue shift of $-$0.03 for an increase in
$\alpha$/Fe by a factor of four.  A solar metallicity population shifted by $-$0.07
blueward for the same change in $\alpha$/Fe.  Similar shifts are expected
in $V-K$ with $\Delta(V-K)$ ranging from $-$0.06 for metal-poor populations to $-$0.09
at solar metallicity.

To incorporate these corrections into our models, we assume that $\alpha$/Fe
decreases, in a linear fashion (as Fe increases from SN events), from an initial
value of 0.4 to a solar value (0.0) over one Gyr of time starting two Gyrs after
initial star formation.  Thus, only the first three Gyrs of star formation have
differing $\alpha$/Fe values from solar, these populations already have the low
metallicities at the end of the simulation, which minimizes the effect.  For example,
changes of $\Delta(B-V)=-0.02$ and $\Delta(V-K)=-0.06$ were calculated in a solar
metallicity stellar population at the end of the simulation.

\subsection{BHB Treatment}

Horizontal branch stars are old, low mass ($M < 1 M_{\sun}$) stars which have entered
the helium core burning phase of their lives.  They are bright ($M_V = -5$), of
nearly constant luminosity and range in color from red to blue (RHB and BHB stars).
BHB stars are of interest to galaxy population models for they have similar
characteristics to young stars in color parameter space, although they are not a
signature of recent star formation.

The effect BHB stars on population models is limited in time and metallicity.  BHB
stars are primarily found in metal-poor clusters ($[Fe/H] < -1.4$) and are not found
in any population younger than 5 Gyrs.  Very few BHB stars are found in the solar
neighborhood (Jimenez \etal 1998), presumingly a combination of young age and high
metallicity, so their contribution in field populations is unclear.  At the start of
our constant star formation scenario, the simulation finds 58\% of the stellar
population has the metallicity and age appropriate for a BHB phase.  Following the
prescription of Conroy, Gunn \& White (2009), this corresponds to a $f_{BHB}=0.6$
which translates into a $\Delta(B-V)$ of $-$0.06 and a $\Delta(V-K)$ of $-$0.03 for a
low metallicity SSP ($[Fe/H] < -1.0$).

As each generation is produced with a range of metallicities, fixed by the peak
[Fe/H] given by the chemical enrichment model, the number of stars with $[Fe/H] <
-1.4$ decreases as the mean metallicity of the galaxy increases with time.  For a
galaxy with an ending metallicity near solar, only 3\% of the population will have a
BHB phase.  For a galaxy with a mean [Fe/H]=$-$1.5 up to 30\% of the stellar
population has a BHB phase.  Using Conroy, Gunn \& White's prescription, this results
in a maximal color difference of $\Delta(B-V) = -0.03$ and $\Delta(V-K) = -0.02$ at
low metallicities, decreasing to zero at solar metallicities.

\subsection{Blue Stragglers}

Blue stragglers stars (BSs) occupy a position in the HR diagram that is slightly bluer
and more luminous than the stellar populations main sequence turnoff point
(Sarajedini 2007).  Their extended main sequence lifetime appears to be due to binary
mass exchange, either by close contact binaries (McCrea 1964) or direct collision
(Bailyn 1995).  Their importance to stellar population synthesis is that they occupy
a region of the HR diagram that mimics star formation and low metallicity effects
(i.e., increased contribution to the blue portion of the integrated SED).  

The more general problem of binary star evolution is outlined in Li \& Han (2008)
which takes into account binary interactions such as mass transfer, mass accretion,
common-envelope evolution, collisions, supernova kicks, angular momentum loss
mechanism, and tidal interactions (Hurley, Tout \& Pols 2002).  The results from
those simulations indicate that, while BSs's are difficult to model and relatively
time sensitive, they are similar to BHB stars in that they only contribute after a
well-formed turnoff point develops at 5 Gyrs.  If collisions are important for their
development, then they will be more rare in galaxy stellar population due to lower
stellar densities compared to globular clusters.  They appear to be numerous in the
Milky Way field populations (Preston \& Sneden 2000), but are an order of magnitude
less luminous than BHB stars.

The simulations of Li \& Han (2008) display a maximum of $-$0.03 bluer colors in
$B-V$ and $-$0.10 bluer colors in $V-K$ for populations older than 5 Gyrs.  The
spread in metallicity is small for $B-V$, approximately 0.01 and non-existent for
$V-K$.  For older populations, it appears that BHB stars dominate over BSs stars
simply based on the fact that BHB corrections to SSP and elliptical narrow band
colors are sufficient to reproduce the color-magnitude relation (Schombert \& Rakos
2009).  For our scenario of constant star formation, by the epoch where BHB stars
decrease in their contribution ($\tau < 5$ Gyrs), BSs stars would beginning to
influence the bluest wavelengths. However, young stars quickly overwhelm the BSs
luminosities and our simulations indicate that the BSs contribution is negligible
when compare to other factors.

\subsection{TP-AGB Treatment}

Most important for our near-IR colors is the treatment of thermally-pulsating AGB
stars (TP-AGBs).  These are stars in the very late stages of their evolution powered
by a helium burning shell which is highly unstable.  They are stars with high initial
masses ($M > 5 M_{\sun}$) and intermediate in age ($\tau > 10^8$ years).  While the
Bruzual \& Charlot codes used for our simulations include TP-AGBs as part of their
evolutionary sequence, comparison with other codes (e.g., Maraston 2005) finds
discrepancies in the amount of luminosity from this short lived population.

\begin{figure}[!ht]
\centering
\includegraphics[scale=0.80,angle=0]{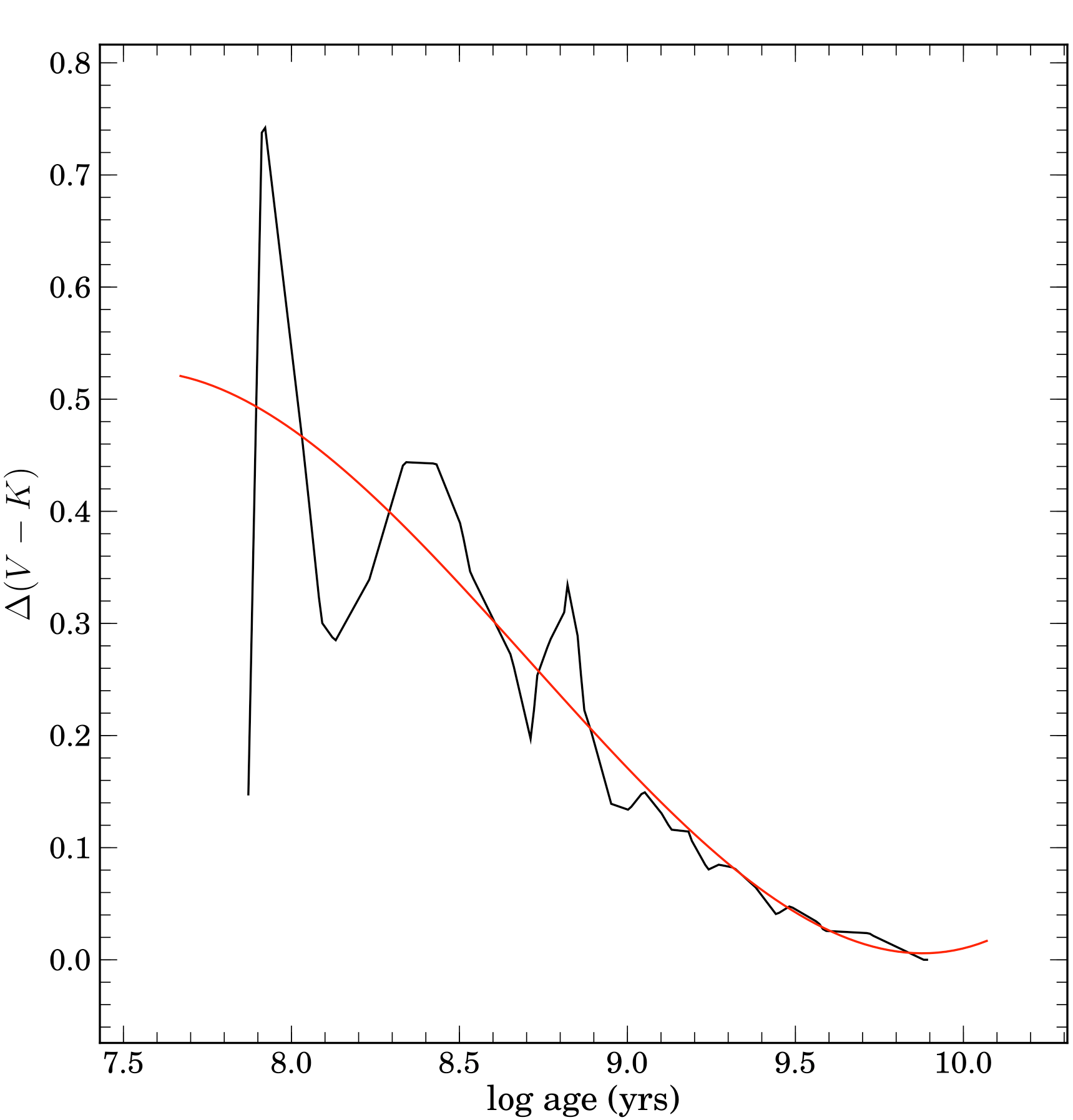}
\caption{\small The black curve is the corrections to BC03 $V-K$ colors from Marigo
\etal (2008), adopted from Maraston (2005) and Bruzual \& Charlot (2003).  The red
curve is a 4th order polynomial fit to the model correction for smooth application to
our simulations.  A start time of $5\times10^7$ yrs is assumed for the TP-AGB phase.
}
\label{agb_vk}
\end{figure}

To better account for TP-AGBs luminosities, we have adjusted the BC03 models to
produce additional TP-AGBs light as given by the CB07 code (Bruzual 2007) and
Marigo \etal (2008).  This adjustment is shown for $V-K$ in Figure \ref{agb_vk}.  We
have selected this correction, over others in the literature, due to the fact that
globular cluster $J-K$ and $V-K$ colors, which were poorly fit by BC03 colors
(Schombert \& Rakos 2009) but are now well matched using the CB07 corrections.

The resulting color changes to the constant star formation model is negligible in
optical colors, as expected, and has a magnitude in the near-IR of $\Delta(V-K) =
0.15$.  This is the most significant of the possible stellar population corrections,
particular effecting the younger populations in the simulation where the optical
colors were assumed to dominate.  This also has a large impact on $M/L$ estimates
made from near-IR luminosities (see \S 4.3).

\section{S$^4$G Comparison Sample and 3.6$\mu$m Colors}

Part of the motivation for these models is for comparison to new {\it Spitzer}
3.6$\mu$m photometry of LSB galaxies (Schombert \& McGaugh 2014).  However, the
spectral libraries do not extend to 3.6$\mu$m, only having produced magnitudes out to
$K$.  Rather than reproducing all the spectral synthesis work, we have decided to
explore the color behavior between $V-K$ and $K-3.6$ in order to empirically
bootstrap from $K-3.6$ to $V-3.6$.

The largest sample of galaxies with both optical and {\it Spitzer} photometry is the
{\it Spitzer Survey of Stellar Structure in Galaxies} project (S$^4$G, Sheth \etal 2010).
We have combined the S$^4$G photometry with $B$ and $V$ values from the RC3 (de
Vaucouleurs \etal 1991) and $K$ values from 2MASS (Jarrett \etal 2003, corrected for
systematic bias, Schombert 2011).  Culling the sample for galaxies with
optical photometric errors less than 0.10 mags and near-IR errors less than 0.15 mags
produced a final sample of 245 galaxies ranging in $V$ mags from 12.5 to 7.

\begin{figure}[!ht]
\centering
\includegraphics[scale=0.60,angle=0]{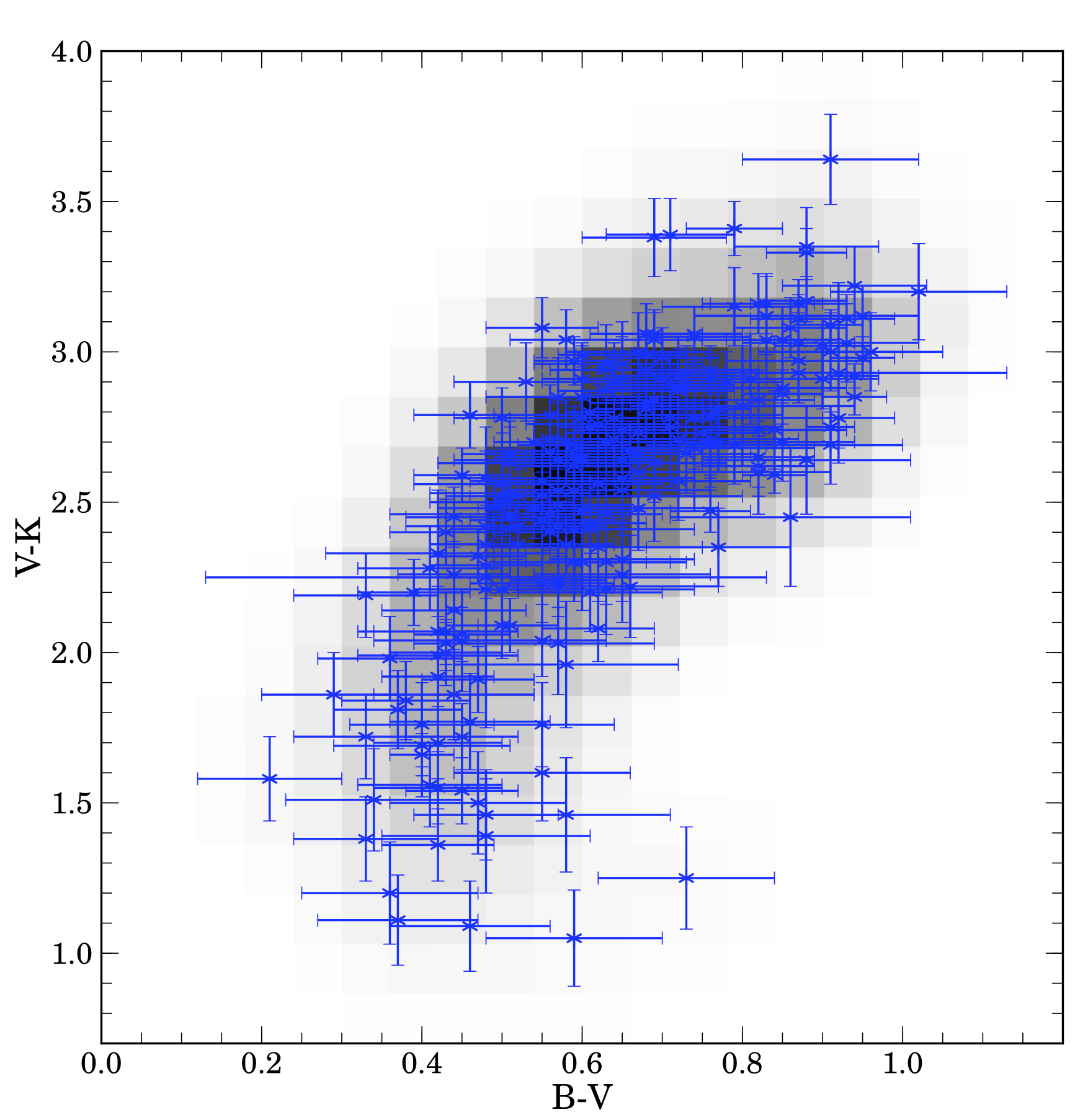}
\caption{\small $B-V$ versus $V-K$ colors for 245 galaxies from the S$^4$G sample (culled
for small photometric errors).  Error bars are shown and the underlying greyscale is
a density plot of the sample.
}
\label{B_V_K_err}
\end{figure}

\begin{figure}[!ht]
\centering
\includegraphics[scale=0.60,angle=0]{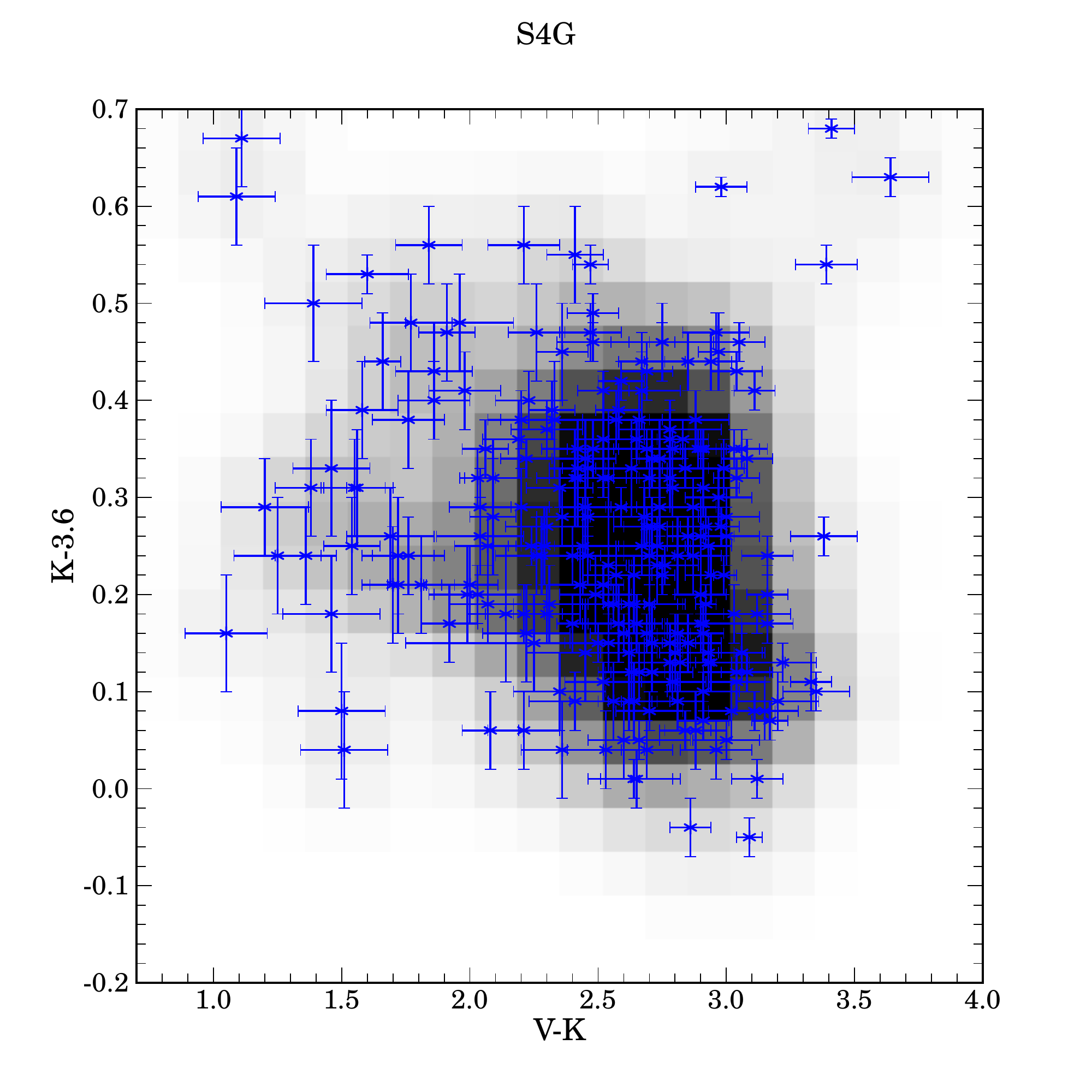}
\caption{\small $V-K$ versus $K-3.6$ colors for 245 galaxies from the S$^4$G sample (culled
for small photometric errors).  Error bars are shown.  In addition, a greyscale
density distribution is also displayed where each data point is assumed to maintain a
2D gaussian with the XY errorbars defining the standard deviation.  The frame is
divided into series of pixels were the greyness of each pixel is based on the sum of
each data points gaussian contribution to provide more definition in the concentrated
regions.
There is no correlation between $V-K$ and $K-3.6$, so
a mean value of $K-3.6=0.27$ is adopted to correct simulation $K$ luminosities into
3.6$\mu$ luminosities.
}
\label{V_K_3.6_err}
\end{figure}

The resulting two color diagrams are shown in Figure \ref{B_V_K_err} and Figure
\ref{V_K_3.6_err} as both density distributions and individual data points with
errorbars.  The trend for bluer $B-V$ for $V-K$ colors is obvious, the curvature
matches the behavior of the stellar population models (see \S 4.1).  The relationship
between $K-3.6$ and $V-K$ is very weak ($R=0.60$) and such that bluer $V-K$ colors
predict redder $K-3.6$ colors.  The large scatter in color below $V-K < 2.0$ is more
than likely due to dust emission contamination from the far-IR by the high SFR members
of the S$^4$G sample.  A better representation of the data is simply that the mean
$K-3.6$ color is 0.27$\pm$0.11 and we will be using this value to correct $V-K$
colors into $V-3.6$ for our models, although for high SFR galaxies a value of $K-3.6$
between 0.4 and 0.5 is more realistic.

\section{Analysis}

\subsection{Photometric Correction Budget}

\begin{figure}[!ht]
\centering
\includegraphics[scale=0.80,angle=0]{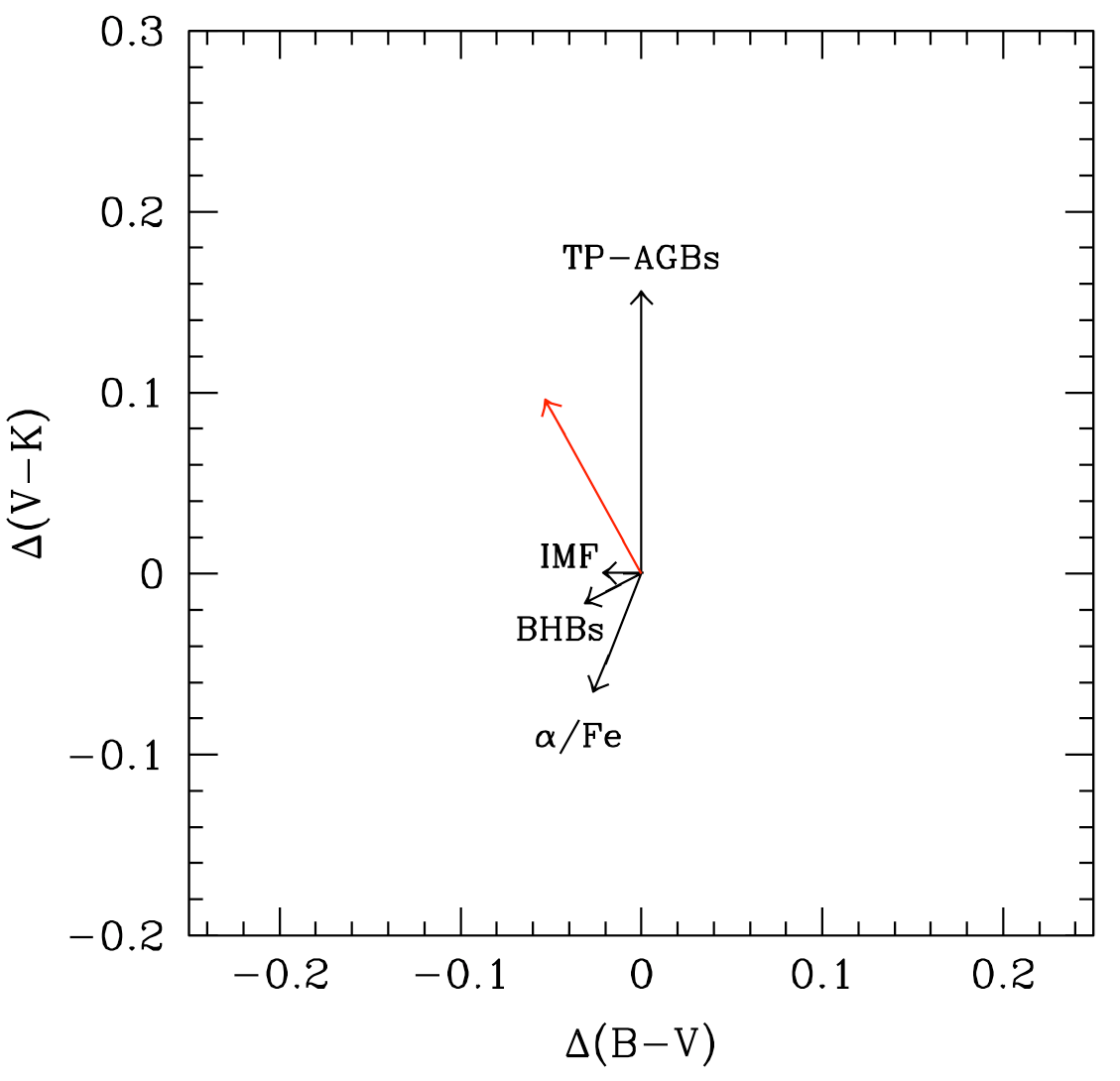}
\caption{\small The various population corrections
outlined in \S4.1.  A majority of the corrections drive galaxy colors
blueward except for TP-AGBs which have a large near-IR redward correction.
The corrections vary slightly with metallicity (becoming smaller with
increasing metallicity).  The adopted mean correction is shown as the red vector.
All these corrections are minor compared to the shifts due to using a
chemical enrichment model and the dominance of young stars.
}
\label{error_budget}
\end{figure}

The effect of the various population corrections as shown in Figure
\ref{error_budget}.  The adopted MDF and chemical enrichment scenarios are assumed
for the simulations and are considered variables to the final model results.  All
population corrections work to drive optical colors blueward, except for the
influence from TP-AGBs which has no effect on blue optical colors.  In the near-IR, the
population corrections for BHBs and $\alpha$/Fe effects are opposite to the
corrections for TP-AGBs (although the TP-AGBs correction is twice the magnitude of
the sum of the other corrections).  The corrections shown in Figure
\ref{error_budget} are adopted from LMC star cluster data, which represents a
population with a mean [Fe/H] of $-$0.5.  Variation in luminosity with metallicity is
approximately a factor of three from metal-poor to solar (Marigo \& Girardi 2007),
but the short lifetimes for TP-AGBs excludes the lowest metallicities from our
models.  The metallicity of the LMC calibrating clusters is a close match the mean
end metallicities of our models, variations in $K$ luminosity up to 30\% can be
expected over the range of model metallicities which translates into a 0.03 variation
in $V-K$ color.  For the purpose of the following discussion, we have adopted the red
vector in Figure \ref{error_budget} as single linear correction to each simulation.
Thus, one should consider the accuracy of the simulations to be limited by about 0.05
in optical colors and 0.07 in near-IR colors.

These corrections, while notable, are small compared to the combined effects of 
chemical evolution model and recent star formation.  As can be seen in Figure
\ref{chemev_color}, by the end of a constant star formation scenario the optical
colors are dominated by the stellar populations from the last 0.5 Gyrs.  The near-IR
colors, on the other hand, would normally track along the optical colors, primarily
influenced by the older metal-poor component from before 5 Gyrs and recent TP-AGBs.

From these preliminary simulations, we know that, while including a multi-metallicity
plus chemical enrichment model is important, small changes in the chemical enrichment
scenario or the shape of the MDF have minor effects on those final colors.  For
example, adopting a gaussian MDF or a linear enrichment scheme only results in 0.005
mag difference in optical or near-IR colors.  Optical colors were sensitive to the
slope of the IMF at the high mass end, but again we have constraints on the upper end slope
based on H$\alpha$ measurements of LSB galaxy star clusters.  The largest change in the model results
from two variables; the star formation start epoch (i.e., the age of the first
generation of stars) and the level of constancy to the star formation rate.  Both to be
discussed in the next sections.

\subsection{Star Formation History Scenarios}

As discussed in \S1, the motivation for a constant star formation model was the
discovery that the SFR for LSB galaxies, based on H$\alpha$ emission, is such that
the ratio of the current SFR to the total stellar mass is similar to a Hubble time.
Star formation must be nearly constant in order to build up sufficient
stellar mass, yet keep the stellar surface densities low.  This is at the heart of
the paradox to the appearance of LSB galaxies, for they require low mean star
formation rates in order to keep their stellar densities low.  However, low star
formation rates produce red optical colors.  The blue nature to LSB also require
sufficient recent star formation to produce hot, young stars.  Thus, a much larger
past SFR, followed by quiescence till the present epoch, is inconsistent with their
present-day optical colors.

\begin{figure}[!ht]
\centering
\includegraphics[scale=0.80,angle=0]{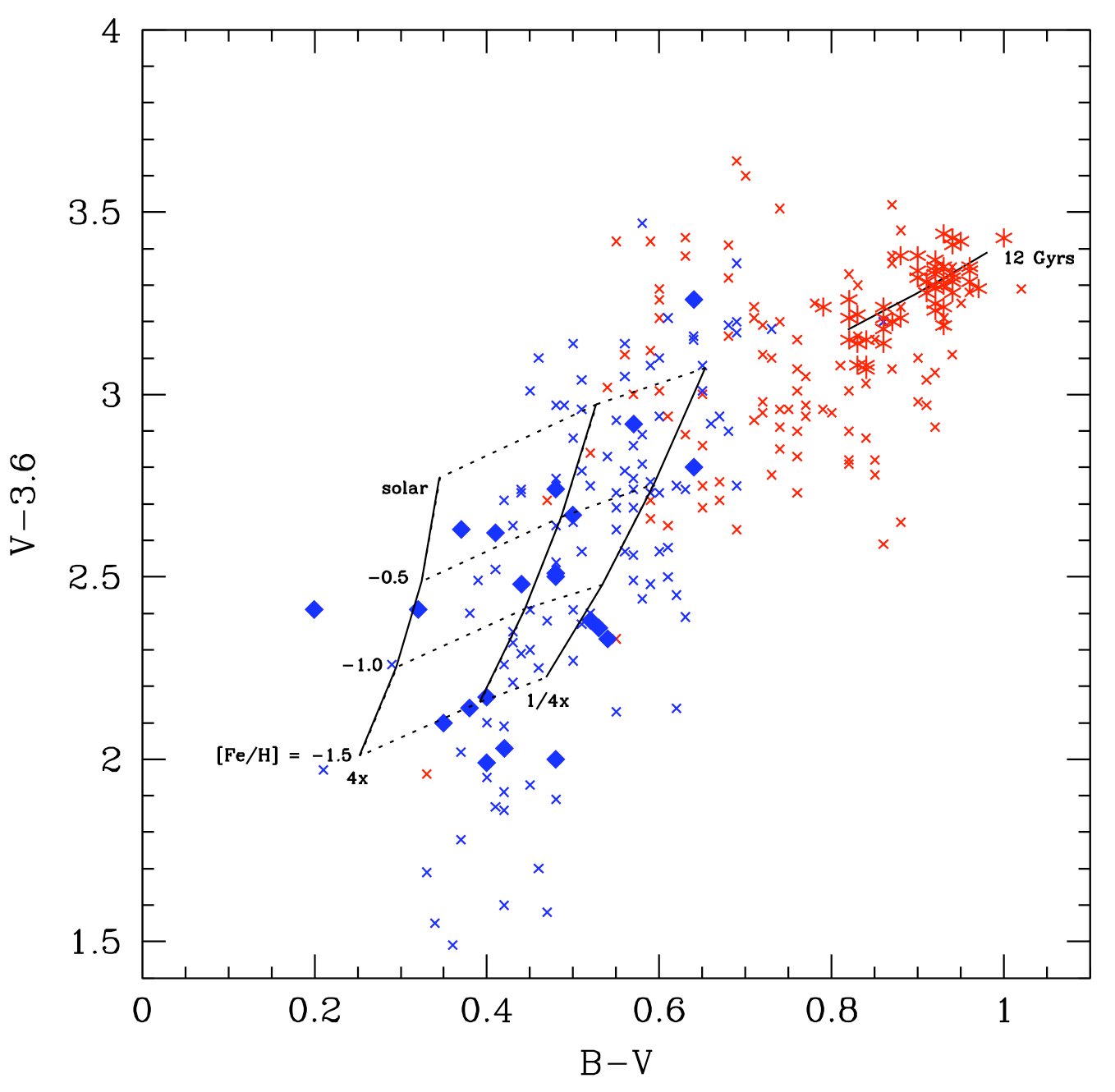}
\caption{\small The two color, $B-V$ vs $V-3.6$ diagram for ellipticals (red stars),
S$^4$G early-type galaxies (red crosses), S$^4$G late-type galaxies (blue crosses)
and LSB galaxies (blue diamonds).  A 12 Gyr multi-metallicity model is shown as a
solid line passing through the elliptical data. The midline track through the S$^4$G
sample represents the constant star 
formation scenario with all the population corrections discussed in the text.  The
simulations ranged from $[Fe/H]= -1.5$ to solar.  Parallel to the constant star
formation track are the low and high burst scenarios where the level of star
formation is increased or decreased by a factor of 4 for the last 0.5 Gyrs of the
simulation.  This covers the range of burst estimates from Lee \etal (2009) and
McQuinn \etal (2010).  A majority of the star forming galaxies are in agreement with
the constant SF model.
}
\label{grid_burst}
\end{figure}

For this reason, all the simulations start times were set to 12 Gyrs (assuming some
small amount of time since the Big Bang for the galaxy to gravitationally collapse).
The simulations were stepped forward in units of 10 Myrs which was shorter than the
lifetime of the most massive star in our stellar tracks.  With a 12 Gyrs formation
epoch, the resulting color tracks for $B-V$ and $V-3.6$, as a function of
metallicity, are shown in Figure \ref{grid_burst} (the midline track).  The displayed
metallicity is the numerical average of the summed stellar populations of all ages,
not a luminosity weighted value.  We note that there are numerous galaxies with
colors redder than our solar endpoint, indicating some super-solar metallicities
exist in star forming galaxies.  In addition, there are a number of galaxies with
very blue $V-3.6$ colors ($V-3.6 < 2$) suggesting that the AGB component in those
galaxies may be overestimated ($V-3.6$ colors of 1.5 are produced by low metallicity
models without any TP-AGBs).

Figure \ref{grid_burst} displays the data for LSB galaxies (Schombert \& McGaugh 2014),
the S$^4$G sample and a set of ellipticals from Dale \etal (2005).  The S$^4$G sample is
divided into early-type (RC3 $T < 4$) and late-type ($T > 4$).  The constant star
formation model is shown as the solid line for various mean [Fe/H].  Also shown is a
12 Gyr SSP, which matches the elliptical data from Dale \etal (2005).  The constant star
formation model matches the mean LSB colors well and agrees with the color relation
for the late-type S$^4$G.  The early-type S$^4$G galaxies lie redward of the models,
indicating the influence of a red bulge on their integrated colors.

It is important to remember that the final colors are independent of the SFR for
a galaxy, as long as the rate is constant.  Low SFR galaxies produce fewer stars than
high SFR galaxies, but the proportion of light each generation is a constant fraction
of the stellar mass.  A high SFR galaxy will have a higher total luminosity (and they
do, see Figure 10 in Schombert, Maciel \& McGaugh 2011), but the colors will be
identical to a lower SFR galaxy.  This explains the lack of any correlation between
color and $L_{H\alpha}$ for many galaxy samples in the literature.  

A scenario for completely constant star formation is only realistic if there is no
variation in the star formation rate.  It should be possible to constrain the
adherence to a constant SF model for LSB galaxies by examining the amount of scatter
(above observational error) in optical colors versus H$\alpha$ flux.  A more direct
measure of the variation in the SFR is the scatter in the $L_{H\alpha}$ as a function
of galaxy mass.  This relationship (see Figure 10, Schombert, Maciel \& McGaugh 2011)
displays a variation by a factor of 4 in $L_{H\alpha}$ (3$\sigma$) scatter from a
linear relation (see also Lee \etal 2009).  As H$\alpha$ emission is a measure of the
SFR over short timescales, we can alter the constant star formation model to
represent quasi-bursts of a factor of 4 intensity over the last 0.5 Gyrs (also shown
in Figure \ref{grid_burst}).  This 0.5 Gyrs timescale for bursts matches the HST
analysis of dwarf galaxy CMD's (McQuinn \etal 2010).  Numerical experimentation
showed that bursts of this magnitude before 0.5 Gyrs fade quickly into the older
population's colors, so the burst only applies to the most recent, and most
metal-rich, epoch.  The application of a burst model to the models can be seen in
Figure \ref{grid_burst}, where the burst models cover most of the spread in LSB
colors, but the late-type S$^4$G sample is better fit by the low burst (1/4x)
scenario than the high burst (4x) one.

The constant star formation scenario for LSB galaxies was suggested by Figure 7 of
Bell \& de Jong (2000), where there is a strong decrease in mean stellar population
age with fainter surface brightness.   Their interpretation is that stellar surface
density maps into gas surface density such that lower LSB galaxies have lower SFR due
to low gas densities.  This also produces less chemical evolution and, thereby, low
mean metallicities.  Our models propose the same results, although we make no claim
on the mechanism behind our scenarios.

The burst scenario is in agreement with the FUV results of Boissier \etal (2008).
The FUV colors are most sensitive to recent star formation, and Boissier \etal find
LSB galaxies have lower SFR than normal spirals (see also our H$\alpha$ observations
from Paper I).  They also interpret the range in FUV-NUV colors of LSB galaxies as
burst and quenching of recent star formation (on timescales of 100 Myrs).  And, they
also conclude that star formation is relatively constant on timescales of Gyrs, as
our models also deduce.  Thus, the important of UV and near-IR colors to map the SFH
of LSB galaxies can not be overstated.

\begin{figure}[!ht]
\centering
\includegraphics[scale=0.80,angle=0]{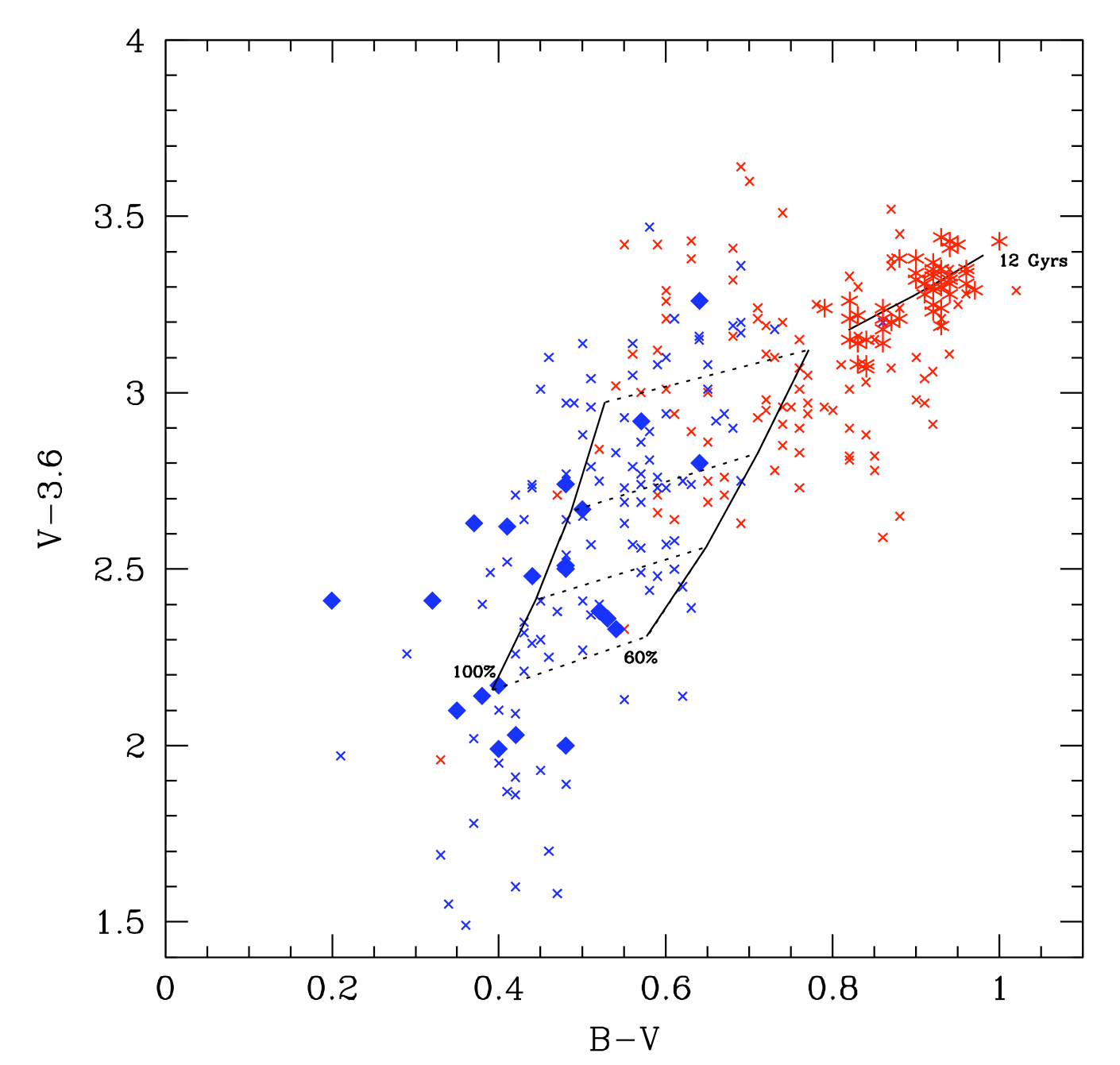}
\caption{\small 
The same as Figure \ref{grid_burst} where the displayed models
represent a declining star formation scenario where the SFR is decreased in a linear
fashion to 60\% of the initial star formation rate.
The scenario accounts for a majority of the early-type galaxies in the S$^4$G sample.}
\label{grid_decline}
\end{figure}

The median color line for the early-type S$^4$G sample suggests that the current star
formation rate is less than the past rate.  This is in agreement with the birthrate
estimates based on their gas supply, where the ratio of gas to stellar mass indicates
more star formation in the past (McGaugh \& de Blok 1997).  To consider a scenario of
declining star formation, we have take a mean birthrate function of 75\% and adjusted
the star formation to decline linearly from 12 Gyrs to the present.  To match the
birthrate function, this requires a change in the current star formation rate of 60\%
from the initial rate.  This differs slightly from exponential star formation
scenarios (Bell \& de Jong 2000), but achieves approximately the same population at
the end of the model.  The resulting track is shown in Figure \ref{grid_decline}
using the same notation as Figure \ref{grid_burst}.  The declining star formation
model brackets the late-type S$^4$G sample and explains a majority of the early-type
spirals as well.

Lastly, we consider a different star formation start epoch, other than the assumed 12
Gyrs for the previous simulations.  This is an old hypothesis that LSB galaxies are
actually younger than other spirals or irregulars (McGaugh, Schombert \& Bothun
1995).  To this end, we simply changed the start time for the simulations in
increments of Gyrs from 12 to 5 Gyrs.  The results were remarkable consistent
in that there is some change in $V-3.6$ and a majority of the shift is in $B-V$.  The
color change was linear until approximately 6 Gyrs with a shift of $\Delta(B-V) =
-0.01$ per Gyrs and a shift of $\Delta(V-3.6) = -0.005$ per Gyr.  This rapidly rules
out any normal or HSB spiral as being less than 12 Gyrs in age, a result known for
decades (Tully \etal 1982).  However, populations of ages between 10 and 12 Gyrs, for
S$^4$G and LSB galaxies, can not be ruled out by the data.

\subsection{Mass-to-Light Ratios}

As noted in McGaugh \& Schombert (2014), many commonly employed models fail to
provide self-consistent results with respect to the $M/L$ values in galaxies.  The
stellar mass estimated from the luminosity in one band can differ grossly from that
of another band for the same galaxy. Independent models agree closely in the optical,
but diverge at longer wavelengths.  

\begin{figure}[!ht]
\centering
\includegraphics[scale=0.80,angle=0]{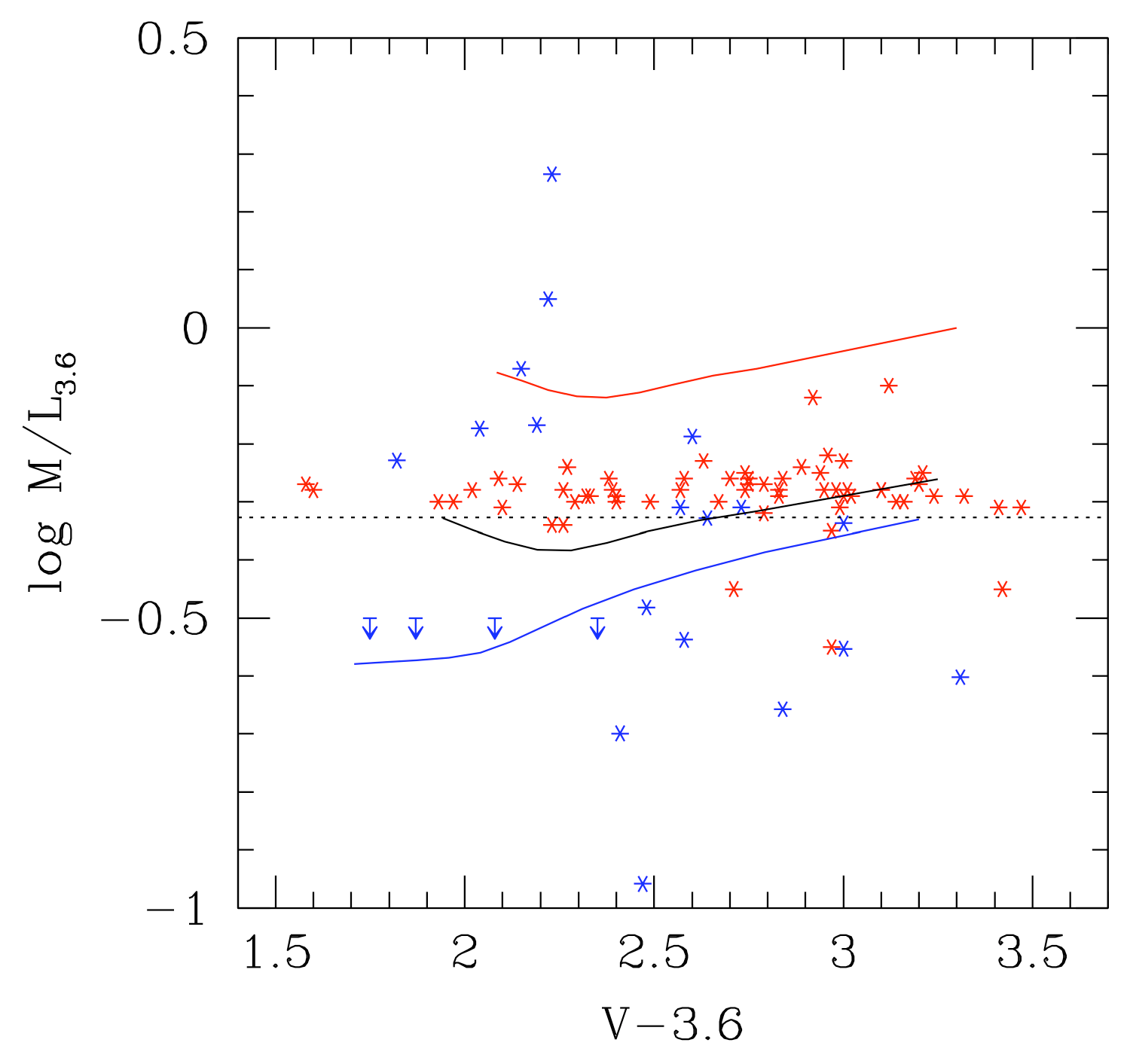}
\caption{\small The relationship between $M/L_{3.6}$ and galaxy color ($V-3.6$).  The
models for constant SF (black line, extended to redder colors), declining SF (red
line) and a burst model (blue line).  The bluest colors represent the lowest
metallicities ([Fe/H]=$-$1.5) and solar metallicity is approximately $V-3.6=3.0$.  In
comparison, the $M/L$ data from Zaritsky \etal (2014) is shown (red symbols) and
$M/L$ estimates from the baryonic TF relation (blue symbols, McGaugh \& Schombert
2014).  The mean $M/L$ estimate of 0.45 from McGaugh \& Schombert is shown as the
dotted line.
}
\label{ml}
\end{figure}

The tension in $M/L$ values from population models is due to the competing
contributions of mass from numerous, older, low mass stars and the high luminosity
from new, hotter stars.  The style of star formation is critical to the evolution of
$M/L$ as recent star formation reduces the contribution from older stars, and
increasing the flux from high mass stars resulting in a lower observed value for
$M/L$.

For our simulations, the shape of the lower end of the IMF effectively ties the mass of the
populations for the upper mass end of the IMF, which contains the bluest stars, are
a negligible component to the total mass of the population.  While we have good
constraints on the upper end of the IMF for LSB galaxies, from the correlation
between optical luminosity and $L_{H\alpha}$, the colors of LSB galaxies are
dominated by both high mass and evolved stars, and do not give us constraints on the low
mass end of the IMF.  While we have no reason to believe that the IMF of LSB
galaxies differs from the standard IMF at the low mass end, this will be the largest
uncertainty in our $M/L$ predictions.

The driving factors for the luminosity in $M/L$ will be wavelength dependent.  For
the optical colors, luminosity in $M/L$ is dominated by very blue, short-lived high
mass stars.  For near-IR color, the luminosity component is sensitive to TP-AGBs,
which have longer lifetimes than high mass stars, but whose contribution can vary
with small increases/decreases in recent star formation (i.e., bursts).  Here we can
be guided by the match between the models and the galaxy colors in Figures \ref{grid_burst}
and \ref{grid_decline} which is quite good.

The resulting $M/L$ tracks for our constant, burst and declining SF models (as
discussed in \S4.2) are shown in Figure \ref{ml} as a function of $V-3.6$ color.  Here
the bluest colors are the lowest metallicity models ([Fe/H] = $-$1.5) and the reddest
colors are models with super-solar metallicity.  As
suspected from early models (Bell \& de Jong 2001; Bell \etal 2003; Portinari \etal
2004; Zibetti \etal 2009; Into \& Portinari 2013), the change in $M/L$ with
metallicity is effectively zero for far-IR colors such that if the star formation
rate is known, the resulting $M/L$ is a constant value for a range of total
metallicities (i.e., galaxy mass).  However, the zeropoint for each model varies from
scenario to scenario.  This is also expected since the primary variable for each
model is the gas consumption rate.  Models loaded with early star formation produce
high $M/L$ values and $M/L$ decreases as the star formation becomes more constant with
time.  The low and high burst models also capture the same behavior since the value
of $M_*$ is constant at late epochs, but the value of $L$ increases or decreases
sharply from enhanced or depressed star formation in the last 0.5 Gyrs.

To demonstrate the importance of determining which star formation scenario is
appropriate, we have plotted the S$^4$G sample $M/L$ estimates from Zaritsky \etal
(2014) using the prescription of Eskew, Zaritsky \& Meidt (2012).  Here the $M/L$
values are determined empirically from the relation between stellar mass to $L_{3.6}$
in the LMC (extrapolating by a factor of 1000 from clusters with $10^7 M_{\sun}$ to
galaxy masses).  The range of $M/L$ is limited due to the lack of strong correlation
between cluster mass and 3.6 or 4.5$\mu$m colors, thus the stellar mass is simply a
linear function of 3.6 luminosity resulting in a constant $M/L$ of $0.52\pm0.07$.  The
constant SF model produces a range of $M/L$ from 0.41 to 0.55 with a mean value of
0.45.  The average $M/L$ from the S$^4$G sample is 20\% higher than the
constant SF model $M/L$, but less than the declining star formation model (mean
$M/L$=0.82), which is a better fit to the S$^4$G colors.

Also shown in Figure \ref{ml} are the $M/L$ estimates from McGaugh \& Schombert
(2014) using the baryonic Tully-Fisher relation to estimate the stellar mass from the
baryonic mass minus the gas mass.  The mean value for that sample is $M/L$ = $0.45\pm0.15$, in
close agreement with the same models that reproduce LSB colors.  As argued in McGaugh \&
Schombert (2014), and supported by the models herein, assuming a constant $M/L$ as
using the 3.6 luminosity of a galaxy results in the most accurate, and repeatable,
stellar mass estimate.

\section{Conclusions}

LSB galaxies represent a unique opportunity in stellar population modeling for their
low current star formation rates mean that their colors are not completely dominated
by the most recent episode of star formation.  In addition, their low metallicities
and stellar densities minimizes complicates due to dust and gas extinction.  Their
gas fractions and stellar mass to SFR ratios limit the range of star formation history
paths to one that is basically constant star formation for the entire life
of the galaxy.  

We summarize our model results as the following:

\begin{description}

\item{(1)} The primary difference between our current set of stellar population
models and previous work is 1) using a full multi-metallicity treatment of the
underlying population, 2) adopting a realistic chemical evolution model and IMF and
3) including corrections for $\alpha$/Fe, BHBs, BSs and TP-AGBs effects.  The error
budget is displayed in Figure \ref{error_budget}, only TP-AGB stars have a
significant component, and only in the near-IR colors.  Using {\it Spitzer}
observations, we can extend the model colors to 3.6$\mu$m from correlations between
$K$ and 3.6.

\item{(2)} As demonstrated with ellipticals (Rakos \& Schombert 2009),
simple stellar populations (SSP's) of a single metallicity are inadequate
to describe galaxy colors.  Applying a metallicity distribution function to
a range of SSP's allows one to construct a multi-metallicity composite
population that merges the colors and luminosities of a range of stellar
metallicities at a particular age.

\item{(3)} The present day colors of LSB galaxies can be modeled by a constant star
formation scenario where the time of initial star formation is assumed to be 12 Gyrs.
Each generation is assigned a mean metallicity to its MDF by following a standard
chemical evolution prescription.  The only free parameter is the final [Fe/H] value
of this chemical evolution model, basically a proxy for the mean SFR as it reflects
into the mean SN rate.  The resulting track as a function of total galaxy metallicity
is shown in Figure \ref{grid_burst}.

\item{(4)} Due to the constant star formation aspect of our models, only
the shape of the upper IMF contributes to present-day colors and
luminosities.  We have solid constrains on the upper end of the IMF in LSB
galaxies based on the correlation between star cluster luminosity and
H$\alpha$ fluxes (Schombert, McGaugh \& Maciel 2013).  We note that uncertainties
in the lower mass end of the IMF will only contribute to the uncertainty in model
$M/L$ estimates.

\item{(5)} Two other SF scenarios are considered, quasi-bursts and a declining 
star formation.  The burst scenarios are where the star formation proceeds in a
quasi-constant fashion with increases/decreases of star formation of a factor of four
over 0.5 Gyrs timescales.  This represents the range in scatter for the galaxy
luminosity versus $L_{h\alpha}$ relation.  The second scenario to let the initial
star formation to be significantly higher than the current SFR, then allow the SFR to
decrease in a linear fashion to today's rate.  This overproduces the amount of
stellar mass in older stars and reddens colors.

\item{(6)} The early-type spirals in the S$^4$G sample are not well fit by the
constant star formation model (being too red in optical colors).  A declining star
formation model, were the current star formation is 60\% of the past average, is
adequate to explain their optical and near-IR colors, although a range of percentages
can be fit to the individual data points.  None of the scenarios match S$^4$G or LSB
colors is the formation epoch is decreased by more than 2 Gyrs.

\item{(7)} The constant star formation model predicts constant $M/L_{3.6}$ in
agreement with previous results using the baryonic Tully-Fisher relations (McGaugh \&
Schombert 2014).  The model and baryonic TF mean $M/L$ is 0.45, which is only 20\%
lower than the mean $M/L$ from LMC clusters given the large statistical uncertainties
(Eskew, Zaritsky \& Meidt 2012).

\end{description}

The motivation for this set of models is the high gas fractions in LSB galaxies and
should have limited application for HSB galaxies or early-type morphologies.  Figure
\ref{grid_burst} would support this conclusion as the colors of most of the S$^4$G
sample are redder than the models.  The models do predict the colors of LSB galaxies,
and a substantial fraction of the late-type S$^4$G galaxies.  Small modifications to
the constant star formation assumption, for example a slightly declining star formation
rate, recovers the most of the colors of early-type spirals, but without uniqueness
to the predicted SF history.

These models illuminate many of the paradoxical features to LSB galaxies.  Low
current star formation rates are in agreement with their low stellar densities.
Their blue optical colors can be reconciled with low stellar densities if the star
formation has been constant or in a weak burst fashion.  The models rule out the
conclusion that LSB galaxies are ``young", in the sense that their epoch of initial
star formation is recent, requiring ages of at least 10 Gyrs.  The models tolerate a
difference in formation age of 2 Gyrs between HSB and LSB galaxy population, but
strongly rule out any difference greater than 5 Gyrs.  Additional understanding of
the stellar populations in LSB galaxies will required high resolution imaging of a
few nearby LSB galaxies with HST, the topic of the fifth paper in our series.

\acknowledgements Software for this project was developed under NASA's AIRS and ADP
Programs. This work is based in part on observations made with the Spitzer Space
Telescope, which is operated by the Jet Propulsion Laboratory, California Institute
of Technology under a contract with NASA.  Support for this work was provided by NASA
through an award issued by JPL/Caltech. Other aspects of this work were supported in
part by NASA ADAP grant NNX11AF89G and NSF grant AST 0908370. This research has
made use of the NASA/IPAC Extragalactic Database (NED) which is operated by the Jet
Propulsion Laboratory, California Institute of Technology, under contract with
the National Aeronautics and Space Administration.

\end{document}